\documentstyle[12pt]{article}

\begin{document}

\newcommand{\Om}{\Omega}
\newcommand{\df}{\stackrel{\rm def}{=}}
\newcommand{\co}{{\scriptstyle \circ}}
\newcommand{\de}{\delta}
\newcommand{\lb}{\lbrack}
\newcommand{\rb}{\rbrack}
\newcommand{\rn}[1]{\romannumeral #1}
\newcommand{\msc}[1]{\mbox{\scriptsize #1}}
\newcommand{\dsp}{\displaystyle}
\newcommand{\scs}[1]{{\scriptstyle #1}}

\newcommand{\ket}[1]{| #1 \rangle}
\newcommand{\bra}[1]{| #1 \langle}
\newcommand{\vac}{| \mbox{vac} \rangle }

\newcommand{\e}{\mbox{{\bf e}}}
\newcommand{\va}{\mbox{{\bf a}}}
\newcommand{\bc}{\mbox{{\bf C}}}
\newcommand{\br}{\mbox{{\bf R}}}
\newcommand{\bz}{\mbox{{\bf Z}}}
\newcommand{\bq}{\mbox{{\bf Q}}}
\newcommand{\bn}{\mbox{{\bf N}}}
\newcommand {\eqn}[1]{(\ref{#1})}

\newcommand{\cp}{\mbox{{\bf P}}^1}
\newcommand{\n}{\mbox{{\bf n}}}
\newcommand{\sbz}{\msc{{\bf Z}}}
\newcommand{\sn}{\msc{{\bf n}}}

\newcommand{\be}{\begin{equation}}\newcommand{\ee}{\end{equation}}
\newcommand{\bea}{\begin{eqnarray}} \newcommand{\eea}{\end{eqnarray}}
\newcommand{\ba}[1]{\begin{array}{#1}} \newcommand{\ea}{\end{array}}

\newcommand{\cleqn}{\setcounter{equation}{0}}
\makeatletter
\@addtoreset{equation}{section}
\def\theequation{\thesection.\arabic{equation}}
\makeatother

\def\np{Nucl. Phys. {\bf B}}
\def\pl{Phys. Lett. {\bf B}}
\def\mpl{Mod. Phys. {\bf A}}
\def\ijmp{Int. J. Mod. Phys. {\bf A}}
\def\cmp{Comm. Math. Phys.}
\def\prd{Phys. Rev. {\bf D}}

\def\ds{dS_{d_-,d_+}}
\def\ads{AdS_{d_-,d_+}}
\def\min{{\cal M}_{d_- +1,d_+}}
\def\ds{dS_{d_-,d_+}}
\def\g{G}

\def\va{\vec a}
\def\vb{\vec b}
\def\vu{\vec u}
\def\vv{\vec v}
\def\vt{\vec t}
\def\vn{\vec n}
\def\ve{\vec e}
\def\vx{{\vec x}}
\def\vxM{{\vec x}_{+}}
\def\vxm{{\vec x}_{-}}
\def\vwM{{\vec w}_{+}}
\def\vwm{{\vec w}_{-}}
\def\vnM{{\check n}_{+}}
\def\vnm{{\check n}_{-}}
\def\dM{{d_{+}}}
\def\dm{{d_{-}}}
\def\ro{r_{0}{}}
\def\vS{\vec {S}}
\def\vsuno{{{\vec s}_1}}
\def\vsdos{{{\vec s}_2}}
\def\ym{y}
\def\hX{{\hat X}}
\def\hJ{{\hat J}}
\def\hP{{\hat P}}
\def\hK{{\hat K}}
\def\hD{{\hat D}}

\newcommand{\matriz}[4]{\left(
\begin{array}{cc}#1&#2\\#3&#4\end{array}\right)}

\begin{flushright}
La Plata Th-99/04\\April, 1999
\end{flushright}

\bigskip

\begin{center}

{\Large\bf Non linear realizations of isometry groups, conformal algebras and geodesics
in Anti-de Sitter like spaces}
\footnote{This work was partially supported by CONICET, Argentina.}

\bigskip
\bigskip

{\it \large Adri\'{a}n R. Lugo} \\ {\sf
lugo@dartagnan.fisica.unlp.edu.ar}
\bigskip

{\it Departmento de F\'\i sica, Facultad de Ciencias Exactas \\
Universidad Nacional de La Plata\\ C.C. 67, (1900) La Plata,
Argentina}
\bigskip
\bigskip

\end{center}

\begin{abstract}

We present the explicit global realization of the isometries of anti-de Sitter like
spaces of signature $(d_-,d_+)$, and their algebras in the space of functions
on the pseudo-Riemannian symmetric space $SO(d_- +1,d_+)/SO(d_-,d_+)$.
The process of going to the invariant boundaries leads to the realization of the
corresponding conformal groups and algebras.
We compute systematically the geodesics in these spaces by considering the coset
representation of them.

\end{abstract}

\bigskip
\section{Introduction}
\cleqn

Anti de Sitter spaces (AdS) were left aside initially in General Relativity in detriment of
de Sitter (dS) spaces due to the (dubious) interpretation of the negative
cosmological constant that supports them.
A posteriori were relived in the context of Supergravity as possible vacuum
solutions in compactification mechanisms \cite{peter} and the study of black hole
thermodynamics on them \cite{hp}.
Emphasis was certainly put in $(1,3)$ signatures and euclidean
versions in the quantum gravity context.

However during the past year a huge amount of work rounding AdS spaces in different
dimensions was made in relation with Maldacena's conjectures relating on-shell quantum
gravity theories on $AdS_{p+1}$ with off-shell conformal field theories (CFT)
that live on its $p$ dimensional boundary \cite{malda1}.
The reason behind  this relation can be heuristically found in the fact \cite{malda1},
\cite{w1} that any theory of gravity with a vacuum value of the metric tensor given by
$d+1$-dimensional $AdS$ space, having in mind the existence of a time-like boundary, should
depend on boundary conditions defined on the $d$ dimensional boundary manifold and should
realize the isometry group of this space that becomes automatically the conformal group
acting on the boundary.
So it seems natural that the degrees of freedom living on the boundary should realize
this symmetry.
Of course things are not so easy, the growing number of people working on the field
extended in many directions a representation of this fact, neither obvious \cite{ps}.

It is the aim of this paper to clarify several aspects of this relation from
the purely mathematical point of view, particularly the group actions on this
spaces and their relation with the corresponding ones on the boundaries, and to
study geodesics on them, all in the general case of $(d_-,d_+)$ arbitrary signature
(since now on $\ads$), generalizing in some cases results known in cases
of earlier physical interest \cite{he}.

\section{Review of $\ads$ spaces}
\cleqn

Several definitions of $\ads$ spaces are possible (in Section 5 we will introduce
another different to that given here useful for our purposes there) and many
parametrizations of them were introduced in the literature, each one useful in particular
contexts.
For sake of completeness and later use we believe pertinent to give a brief summary of them.

\subsection{The defining coordinates}

Let us consider cartesian coordinates $\;\{ w^M , M = -d_-, \ldots,0,\ldots,d_+ \}\;$ in
$\Re^{1+ d_- + d_+}$ with $\;d_\pm \geq 0,$
and endows it with the minkowskian metric
\be
\eta_{d_- +1,d_+} = \eta_{MN}\; dw^M\;  dw^N\;\; , \;
\eta \equiv \matriz{-1_{d_- + 1}}{0}{0}{1_{d_+}}\label{gmin}
\ee
This is a maximally symmetric space (MSS) with the pseudo-Poincar\`e isometry
group $ISO(d_- + 1,d_+)$.
The $\ads$ space is introduced as the submanifold defined by the constraint
\be
\eta_{MN}\; w^M\;  w^N = -\ro^2 \;\;, \;\; \ro>0\label{cons}
\ee
where $\ro$ is the radius (curvature scale) of the space, with the
metric $G$ induced by \eqn{gmin}.
Clearly the constraint respects only $SO(d_- + 1, d_+)$ invariance and then it
becomes the isometry group of $\ads$; reduction of the dimensionality combines with
the reduction of symmetry to make it also a MSS.
As such its curvature tensor verifies
\footnote{We adopt the conventions of \cite{wald}; also the convention of rising and lowering
indices with the corresponding (from the context) $\eta$-matrix is implicit all along the paper.
}
\bea
{\cal R}_{abcd} &=& \ro^{-2}\; (  G_{ad}\; G_{bc} - G_{ac}\; G_{bd} )\cr
{\cal R}_{ab} &=& \frac{2}{\dm +\dM -2}\; \lambda\; G_{ab}\cr
{\cal R} &=& -\;\ro^{-2}\; (d_- + d_+ -1 ) \; (d_- + d_+ )
\eea
where $\; \lambda = -\; (2\ro^2 )^{-1}\; (d_- + d_+ -1 )\;(d_- + d_+ -2 )\;$ is the
cosmological constant.
In this defining coordinates we can solve for one of the coordinates, i.e.
\be
w^{-d_-} = \pm (w^\mu w_\mu + \ro^2 )^{\frac{1}{2}} \label{w-}
\ee
and parametrize (each sheet of) $\ads$ by
\be
\{ w^\mu , \;\;\mu = -d_- +1, \ldots,0,\ldots,d_+\} = \left(\begin{array}{c}
 \vwm \\ \vwM  \end{array}\right)\;\in\Re^{d_-}\times\Re^{d_+}\label{dg}
\ee
The metric takes the form
\be
\g = \left(\eta_{\mu\nu} - \frac{w_\mu\; w_\nu}{w^\rho\;w_\rho +
\ro^2}\right)\; dw^\mu\; dw^\nu\label{gdc}
\ee
In the euclidean case $d_-=0$ the topology corresponds to a two-sheeted hyperboloid with
basis in $w^{0} = \pm\ro $; we will consider the connected component given by the $+$
sign in $\eqn{w-}$, the sheet based on $w^0 = +\ro$ with naive topology $ \Re^{d_+}$.
If instead $d_+ = 0$ the topology is compact corresponding to a $S^{d_-}$ of radius
$\ro$ and ``wrong" sign metric.
In the general case $d_\pm >0$ the topology is $S^{d_-}\times \Re^{d_+}$
corresponding to a hyperboloid, the (non) compact directions being
(space) time like.
It is important to note that the solution of \eqn{w-} imposes the
condition
\be
w^\mu \; w_\mu + \ro^2 > 0 \label{conboundc}
\ee

\subsection{The plane coordinates}

They are defined by the following one-to-one map
\be
z^\mu = \frac{w^\mu}{1+ (1+\frac{w^\rho w_\rho}{\ro^2} )^\frac{1}{2}}\;
\longleftrightarrow\; w^\mu = \frac{2\; z^\mu}{1 - \frac{z^\rho z_\rho}{\ro^2}} \label{plc}
\ee
In terms of them the metric takes the simple diagonal form where the signature is
manifest,
\be
\g = \frac{4}{(1 - \frac{z^\rho z_\rho}{\ro^2} )^2} \;
\eta_{\mu\nu}\; dz^\mu\; dz^\nu
\ee
The definition \eqn{conboundc} togheter with \eqn{plc} is then equivalent to
\be
-\ro^2 < z^\mu \; z_\mu  < \ro^2  \label{conbounpc}
\ee
Let us remember that unless $\dm =0$ these coordinates cover only
one-half of $\ads$; in fact from the relations
\be
z^\rho\; z_\rho = \frac{w^\rho\; w_\rho}{\left( 1 +
(1 + \frac{w^\rho\; w_\rho}{\ro^2})^{\frac{1}{2}}\right)^2 }\;\; \longleftrightarrow \;\;
w^\rho\; w_\rho = \frac{4\;z^\rho\; z_\rho}{\left( 1 - \frac{z^\rho\; z_\rho}{\ro^2} \right)^2}
\ee
we can have
\be
w^{-\dm} = \pm\;\ro\;\frac{\ro^2 + z^\rho\; z_\rho}{\ro^2 -z^\rho\; z_\rho }
\ee
corresponding to the $\pm$ choice in \eqn{w-}.

\subsection{The global coordinates}

Given the constraint \eqn{cons} it is natural to introduce in our context hyperbolic
coordinates covering the sector of $\; w^M w_M < 0\;$ of  $\min$ space as follows,
\bea
\left(\begin{array}{l} w^{-\dm}\\ \vwm \end{array}\right) &=& r\;\cosh r_L\; \vnm \cr
\vwM &=& r\;\sinh r_L\; \vnM   \label{rc}
\eea
where $r>0 , r_L \geq 0,$ and ${{\check n}_\pm}^t{{\check n}_\pm} = 1$,
so $\vnm , \vnM$ define points in $S^\dm, S^{\dM -1}$ respectively (the case $\dM = 0$ is
handled by putting $r_L\equiv 0)$).
Obviously in this coordinates the constraint \eqn{cons} is
equivalent to fix $r = \ro $; the metric is written as
\be
\g = \ro^2\; \left( - \cosh^2 r_L\; d^2 \Omega_{\dm} + d^2 r_L +
\sinh^2 r_L\; d^2 \Omega_{\dM -1} \right)\label{grc}
\ee
where $d^2 \Omega_{p}$ stands for the standard measure of (radius $1$) $S^p$.
The signature as well as the topology is manifest in this form, with $\vnm$ and
$(r_L,\vnM )$ parametrizing $S^{\dm}$ and $\Re^{\dM}$ respectively.
The $``+"$ ($``-"$) sign in \eqn{w-} corresponds to consider the semi-sphere of $S^\dm$
containing the north (south) pole.

It is common in general relativity contexts \cite{hp} to introduce the radial variable
\bea
r_H &=& \sinh r_L \;\;\; , \;\; r_H\geq 0\cr
\g &=& \ro^2\; \left( - (1 + r_H{}^2 ) \; d^2 \Omega_{\dm} + \frac{d^2 r_H}{1 + r_H{}^2 }
+ r_H{}^2\; d^2 \Omega_{\dM -1}\right)\label{ghc}\label{hc}
\eea
where the near horizon geometry $r_H \sim 0$ is manifestly minkowskian.

Another variant is to introduce
\bea
r_H &=& \tan \rho  \;\; ,\;\; 0 \leq \rho <\frac{\pi}{2} \cr
\g &=& \ro^2\; \sec^2\rho\; \left( - d^2 \Omega_{\dm} + d^2\rho + \sin^2\rho\; d^2
\Omega_{\dM -1}\right)\label{grhoc}\label{rhodef}
\eea
which shows that $\ads$ is conformal to the geometry product of a time-like $S^\dm$ and a
``trumpet", the well known (for $(1,3)$ signature) Einstein static universe (for exactness,
one half of it \cite{he}).

\subsection{The Poincar\`e coordinates}

In $\min$ are introduced in terms of $(w^M )$ the following coordinates ($d_+ >0$)
\be
\left\{
\begin{array}{l}  x^\alpha = \ro\; (w^{-\dm} + w^\dM )^{-1}\; w^\alpha \\
x^+ = \ro^2 \;(w^{-\dm} + w^\dM )^{-1} \\
\;V = \ro^{-2} \;(w^{-\dm} - w^\dM )\end{array}\right.
\;\;\longleftrightarrow\;\;
\left\{ \begin{array}{l} w^{-\dm} = \frac{\ro^2}{2}\; ( (x^+)^{-1} + V)\\
\;\;\; w^\alpha = \ro\; (x^+)^{-1}\; x^\alpha\\
\;\; w^{\dM} = \frac{\ro^2}{2}\; ( (x^+)^{-1} - V)\end{array}\right.\label{pc}
\ee
where $\;\alpha = -\dm +1, \ldots, 0, \ldots, \dM -1\;$. The constraint \eqn{cons} looks
\be
-\ro^{2}\; (x^+)^{-1}\; V + (x^+)^{-2}\; \eta_{\alpha\beta}\; x^\alpha\; x^\beta + 1 = 0
\label{conspc}
\ee
and solving for $V$ the metric reads in Poincar\`e parametrization ($ x^+ \equiv x^\dM$)
\be
\g =  \frac{\ro^2}{(x^+)^2} \; \eta_{\mu\nu}\; dx^\mu\;dx^\nu
\label{gwc}
\ee
These coordinates were considered in \cite{w1} to do computations in the boundary theory;
the reason of this fact should be clear in Section 3.

A little modification  of this system of coordinates, namely
\bea
U &=& (x^+ )^{-1}\cr
\g &=& \ro^2 \; \left(\frac{d^2 U}{U^{2}} + U^2\; \eta_{\alpha\beta}\;
dx^\alpha\;dx^\beta\right)\label{gpc}
\eea
was introduced in \cite{malda1} where it appears in a natural way coming from the usual
parametrization of $p$-brane geometries in the decoupling limit.

\subsection{The boundary of $\ads$ spaces}

For $\dM >0$ we saw that $\ads$ is a non compact space.
It is possible however to adscribe to it a boundary through Penrose's concept of conformal
infinity.
The rough idea (see \cite{he} for exact statements) consists in trying to map the space-time
at hand to a (possible finite) region in which null curves are at $\pi/4$, and in such a way
that both geometries are Weyl-related.
Then causality concepts will coincide in both spaces and we should be able to identify
the conformal boundaries where any curve should born and die.

In our case we have the job made by considering the form \eqn{grhoc} of the metric; from there
it follows that the boundary corresponds to the hypersurface defined by $\rho=\pi/2$,
topologically  homeomorphic to $\;S^\dm\times S^{\dM -1}$.
\footnote{
When $\dM = 1$ the boundary consists of two disconnected $S^\dm$ ( $S^0 \sim Z_2$).
}
It is time-like in character; when $\dm =1$ this fact is responsable among other things
for the lack of global hyperbolicity (if $\dm >1$ clearly the concept of a Cauchy surface
is not obvious).

Note also that this definition is the same as to consider the $r_L=\infty$ limiting surface, the
same way in spirit in which the boundary was introduced in \cite{w1}.
In plane coordinates it corresponds to the hypersurface
\be
z^\rho \; z_\rho = r_0^2
\ee
In Poincar\`e coordinates it is sitted at $x^+ = 0$ ($U= \infty$); however some points
``at infinity" should be added to the space parametrized by $(x^\alpha)$ to get the
conformal compactification of ${\cal M}_{\dm,\dM-1}$, i.e. $\;S^\dm\times S^{\dM -1}$.
Of course this fact a bit obscure in this coordinatization follows from the relations
\eqn{pc} and becomes evident from the action of $SO(\dm +1, \dM)$ on the boundary given
in the next section.

\section{The $SO(\dm +1,\dM )$ action}
\cleqn

The action of the isometry group on $\ads$ is that induced by the
linear action on $\min$.
Let $\Lambda_{\dm +1,\dM} \in SO(\dm+1,\dM )$; it admits a coset decomposition under its
maximal compact subgroup $SO(\dm +1)\times SO(\dM)$ in the form \cite{gil}
\begin{eqnarray}
\Lambda_{\dm +1,\dM }(S,P, Q) &=& K_{\dm +1,\dM}(S)\; H(P,Q) \cr
K_{\dm +1,\dM}(S) &=& \matriz{(1 + S S^{t})^{\frac{1}{2}}}{S}{S^{t}}{(1 + S^{t}S)^{\frac{1}{2}}}
= \exp\matriz{0}{N}{N^t}{0} \cr
S &\equiv& \frac{ \sinh (N\; N^t)^\frac{1}{2}}{(N\; N^t)^\frac{1}{2}}\;
N\;\; \in \Re^{(\dm +1 )\times\dM}\cr
H(P,Q ) &=& \matriz{P}{0}{0}{Q} \;\; , \; P\in SO(\dm +1)\; ,\; Q\in SO(\dM) \label{parortpq}
\end{eqnarray}
The linear action on $(w^M)\equiv (w^{-\dm}, \vwm ,\vwM ) \in \min$ is then given by
\bea
\left( \begin{array}{l} {}^\Lambda w^{-\dm} \\ {}^\Lambda\vwm \end{array}\right)
&=&(1 + S S^{t})^{\frac{1}{2}}\; P\;\left( \begin{array}{l} w^{-\dm} \\ \vwm \end{array}\right)
+  S\; Q\; \vwM\cr {}^\Lambda\vwM &=& S^t\; P\; \left( \begin{array}{l} w^{-\dm} \\ \vwm
\end{array}\right) + (1 + S^{t}S)^{\frac{1}{2}}\;
Q\;\vwM\label{lintran}
\eea

By restricting ourselves to the submanifold \eqn{cons} and using
\eqn{w-} we get the non linear realization searched for.
This is easier acomplished in the global coordinates $( r_L , \vnm , \vnM )$ of
\eqn{rc}; in terms of them the transformations are given by
\bea
\sinh {}^\Lambda r_L &=& |\cosh r_L \; S^t\; P\; \vnm + \sinh r_L \;(1 + S^{t} S)^{\frac{1}{2}}\;
Q \; \vnM | \cr
{}^\Lambda\vnm &=&
\frac{\cosh r_L (1 + S S^{t})^{\frac{1}{2}}\; P\;\vnm + \sinh r_L \; S\; Q\; \vnM}
{| \cosh r_L (1 + S S^{t})^{\frac{1}{2}}\; P\;\vnm + \sinh r_L \; S\; Q\; \vnM|}\cr
{}^\Lambda\vnM &=&
\frac{\cosh r_L \; S^t\; P\; \vnm +\sinh r_L (1 + S^{t}S)^{\frac{1}{2}}\; Q\;\vnM }
{|\cosh r_L \; S^t\; P\; \vnm +\sinh r_L (1 + S^{t}S)^{\frac{1}{2}}\; Q\;\vnM |}
\label{nltgc}
\eea
From here it is clear that the boundary results invariant under the $SO(\dm +1,\dM)$
transformations; in fact $r_L = \infty$ is mapped to ${}^\Lambda r_L = \infty$ while the
variables in the boundary manifold realize the isometry in the way
\bea
{}^\Lambda\vnm & \stackrel{r_L=\infty}{=}&
\frac{(1 + S S^{t})^{\frac{1}{2}}\; P\;\vnm + S\; Q\; \vnM}
{|  (1 + S S^{t})^{\frac{1}{2}}\; P\;\vnm + S\; Q\; \vnM|}\cr
{}^\Lambda\vnM &\stackrel{r_L=\infty}{=}&
\frac{ S^t\; P\; \vnm +(1 + S^{t}S)^{\frac{1}{2}}\; Q\;\vnM }
{| S^t\; P\; \vnm +(1 + S^{t}S)^{\frac{1}{2}}\; Q\;\vnM |}
\eea
These expressions provide a global, non linear realization of generalized conformal
transformations.
This realization is made not on ${\cal M}_{\dm,\dM-1}$ but on the conformal compactification of
it, i.e. $S^\dm\times S^{\dM -1} $.
For example in the euclidean case $\dm = 0$ there is no $\vnm$ and the boundary
$S^{\dM -1}$ is parametrized by $\vnM\equiv\vn$; with $S^t\equiv {\vec S}$ we have
\be
{}^\Lambda\vn =
\frac{ Q\; \vn + \left( (1 + |{\vec S}|^2)^{\frac{1}{2}} - 1\right)\; {\check S}^t\; Q\;\vn\;
{\check S} }{\left( 1 + ({\vec S}^t\; Q\;\vn)^2 \right)^\frac{1}{2}}
\ee
The (linear) transformations ${\vec S}= {\vec 0}$ with $Q\in SO(\dM)$ are
just the isometries of $S^{\dM -1}$ and those with $Q=1$ corresponds
to scale and special conformal transformations parametrized by
${\vec S}\in\Re^\dM$; both generate the conformal group $SO(1,\dM )$.
But although compact and clear in meaning they are not enough useful for most current applications.
For this we move to get the realization in Poincar\`e coordinates.

From \eqn{pc} it should be clear that the parametrization \eqn{parortpq} is not a convenient one.
Instead we will proceed in two steps.
First we decompose $SO(\dm +1 ,\dM )$ under $SO(\dm , \dM )$
\bea
\Lambda_{\dm +1,\dM}( \vS , Q_{\dm\dM}) &=&
K_{\dm + 1,\dM}(\vS )\; H(1,Q_{\dm,\dM}) \cr
K_{\dm + 1,\dM}(\vS ) &=& \matriz{C}{ {\vS}^t \; \eta_{\dm\dM} }{\vS}{1_{\dm ,\dM} +
\frac{{\vS} \; {\vS}^t \;\eta_{\dm,\dM}}{1+C} } \label{dec1}
\eea
where $\; Q_{\dm,\dM}\in SO(\dm ,\dM )\;,\; {\vec S} \in \Re^{\dm+\dM}\;$, and
$\; C^2 = 1 + \vS^t \eta_{\dm,\dM}\vS \geq 0\;$ define the coset manifold.
In what follows we will consider separately the $-\dm$ and $\dM$ directions,
$w^{\pm d_\pm } \equiv w^\pm$,  and vectors, $\eta$ matrices, etc.,  will be
($\dm + \dM -1$)-dimensional; scalar products of vectors will be wrt
$\eta_{\dm,\dM -1}\equiv\eta$, i.e.
${\vec v}_1 \cdot {\vec v}_2 \equiv {\vec v}_1{}^t\eta {\vec v}_2$.
In this spirit we introduce
\be
\vS \equiv  \left( \begin{array}{c} \vsuno \\ \sigma \end{array}\right)\;\;\; ,\;\;
C^2 = 1 + \vsuno^2 + \sigma^2 \label{S}
\ee
With this in mind the second step is to decompose $Q_{\dm,\dM}$ under $SO(\dm , \dM -1)$
in a similar fashion
\bea
Q_{\dm ,\dM}( \vsdos , Q) &=&
K_{\dm ,\dM}(\vsdos )\; H(Q, 1) \cr
K_{\dm ,\dM}(\vsdos ) &=& \matriz{1 - \frac{{\vsdos} \; {\vsdos}^t\;\eta}{1+c_2} }
{\vsdos}{-{\vsdos}^{t} \;\eta}{c_2} \label{dec2}
\eea
where $\; c_2{}^2 = 1 - \vsdos^2 \geq 0\;$.
By plugging \eqn{S},\eqn{dec2} in \eqn{dec1} we get another parametrization of $SO(\dm+1,\dM)$
different from \eqn{parortpq} in terms of a $SO(\dm,\dM -1)$ matrix $Q$, two
real $(\dm  +\dM -1)$-dimensional vectors $\vsuno ,\vsdos$ and a real number $\sigma$.

As in \eqn{lintran} is straightforward to get the linear transformations of the
$(w^- , {\vec w} , w^+)$-coordinates.
With them and \eqn{pc}, \eqn{conspc} we finally obtain
\bea
\frac{2 x^+}{ ^\Lambda x^+}\; \frac{^\Lambda\vx}{\ro} &=&
( 1 + \frac{\vsuno\cdot\vsdos + \sigma c_2}{C + 1})\;\vsuno + \vsdos +
\left(( 1 - \frac{\vsuno\cdot\vsdos + \sigma c_2}{C + 1})\;\vsuno -
\vsdos \right)\; \frac{\vx^2 + {x^+}^2}{\ro^2} \cr
&+& 2\; \left[1 + \left( \frac{\vsuno \vsuno^t}{1+ C} - \frac{\vsdos \vsdos^t}{1+
c_2} - (\frac{\vsuno\cdot\vsdos}{c_2 + 1} + \sigma)\;\frac{\vsuno \vsdos^t}{1+ C}
\right)\; \eta\right]\; Q\; \frac{\vx}{\ro}\cr
\frac{2 x^+}{ ^\Lambda x^+} &=&
C + \sigma + c_2 + ( \vsuno\cdot\vsdos + \sigma c_2 )\; (1 +\frac{\sigma}{C+1})\cr
&+& 2\;\left[ (1 + \frac{\sigma}{C+1})\; \vsuno -
\left( (1 + \frac{\sigma}{C+1})\; (\frac{\vsuno\cdot\vsdos}{c_2 + 1} +
\sigma)+ 1 \right)\; \vsdos\right]\cdot Q\frac{\vx}{\ro}\cr
&+& \left(C + \sigma - c_2 - ( \vsuno\cdot\vsdos + \sigma c_2 )\; (1 +
\frac{\sigma}{C+1})\right) \; \frac{\vx^2 + {x^+}^2}{\ro^2}\label{nltpc}
\eea
These are the expressions we liked to find.
They realize non linearly the group of isometries $SO(\dm + 1,\dM)$  on the whole
$\ads$ space, and act on the invariant boundary $x^+ = 0$ (modulo
compactifications) in the usual way.
More explicitly, it is not difficult to show that the standard
conformal transformations are given by

\begin{itemize}
\item\underline{Pseudo-Poincar\`e transformations.}
\be
^{(P,\vec a )}\vx = P\; \vx + \vec a \;\;\longrightarrow\;
\left\{ \begin{array}{l}
Q = P\cr
\vsuno = \ro^{-1}\; \vec a \cr
\vsdos = (1 +\frac{{\vec a}^2}{4\ro^2} )^{-1}\; \frac{\vec a}{\ro}\cr
\sigma = -\frac{{\vec a}^2}{2\ro^2} \end{array}\right.\label{pointr}
\ee
\item\underline{Dilations}
\be
^\lambda\vx = \lambda^{-1}\;\vx
\;\;\longrightarrow\;
\left\{ \begin{array}{l}
Q = 1\cr
\vsuno = \vec 0\cr
\vsdos = \vec 0\cr
\sigma = \frac{1}{2}\; ( \lambda - \frac{1}{\lambda} )
\end{array}\right.\label{diltr}
\ee
\item\underline{Special conformal transformations}
\be
^{\vec b}\vx = \frac{\vx + \vx^2\;{\vec b}}{1 + 2\;{\vec b}\cdot\vx + {\vec b}^2\;\vx^2}
\;\;\longrightarrow\;
\left\{ \begin{array}{l}
Q = 1\cr
\vsuno = \ro\; \vec b\cr
\vsdos = -(1 + \frac{\ro^2}{4}\; {\vec b}^2)^{-1} \;\ro\;\vec b\cr
\sigma = \frac{{\vec b}^2}{2\ro^2}\end{array}\right.\label{spcontr}
\ee
\end{itemize}

\section{The $\ads$ and conformal algebras}
\cleqn

Here we consider the representation of the corresponding algebras and their meaning in
each context.

The $SO(\dm +1, \dM)$ Lie algebra is ($X_{MN} = - X_{NM}$)
\be
[X_{MN}, X_{PQ}] = \eta_{NP}\; X_{MQ} + \eta_{MQ}\; X_{NP} - (M\leftrightarrow N)
\label{stalg}
\ee
An arbitrary element $\delta\Lambda$ in the vector representation obeys:
$\;\delta\Lambda^t\; \eta + \eta\; \delta\Lambda =0\; $, and then
$\;\delta\Lambda = \frac{1}{2} \delta\Lambda^{MN} V_{MN}\;$,
where the generators in this representation are
\bea
V_{MN} &\equiv& V(X_{MN}) = E_{MN} - E_{NM}\cr
(E_M{}^N)^P{}_Q &\equiv& \delta_M^P\; \delta_Q^N
\eea
The infinitesimal generators acting on functions in $\min$ are
introduced in the standard way in order to obey \eqn{stalg}
\bea
\delta_{\delta\Lambda} w^M &\equiv& - \delta\Lambda^{PQ}\; \hX_{PQ} w^M\cr
\hX_{MN} &=& w_M \;\partial_N - w_N\; \partial_M \label{infgen}
\eea
Then the realization of the generators on $\ads$ in $(w^\mu)$ coordinates are gotten by taking
into account the constraint \eqn{cons}
\bea
\hX_{\mu\nu} &=& w_\mu \; \partial_\nu - w_\nu \; \partial_\mu\cr
{\hat p}_\mu &=& \pm \; (w^\rho w_\rho + \ro^2)^\frac{1}{2}\; \partial_\mu = \hX_{\mu -} |_{AdS}
\eea
and obeys the $\ads$ algebra
\bea
[\hX_{\mu\nu}, \hX_{\rho\sigma}] &=& \eta_{\nu\rho}\; \hX_{\mu\sigma} +
\eta_{\mu\sigma}\; \hX_{\nu\rho} - (\mu\leftrightarrow \nu)\cr
[ {\hat p}_\rho ,\hX_{\mu\nu}] &=& \eta_{\mu\rho}\; {\hat p}_\nu -
\eta_{\nu\rho}\; {\hat p}_\mu \cr
[{\hat p}_\mu , {\hat p}_\nu ] &=& \hX_{\mu\nu }\label{infgenads}
\eea
Now let us see the realization in Poincar\`e coordinates.
From \eqn{nltpc} the infinitesimal transformations are
\bea
\delta_{\delta\Lambda} \vx &=&  \delta P\; \vx + \delta\va
- \delta\lambda\; \vx + \delta\vb\; ( \vx^2 + (x^+)^2 ) - 2\;\delta\vb\cdot \vx\;\; \vx\cr
\delta_{\delta\Lambda} x^+ &=& - (\delta\lambda + 2\; \delta\vb\cdot \vx\ )\; x^+
\eea
where in view of \eqn{pointr}, \eqn{diltr}, \eqn{spcontr} we have introduced
\bea
\delta P &=& \delta Q\cr
\delta\va &=& \frac{\ro}{2}\; ( \delta\vsuno + \delta\vsdos )\cr
\delta\lambda &=& \delta\sigma\cr
\delta\vb &=&  \frac{1}{2\ro}\; ( \delta\vsuno - \delta\vsdos )
\eea
and according to the definition in the first line of \eqn{infgen}
we get the infinitesimal generators associated to $\;(\delta P ,
\delta\va , \delta\lambda, \delta\vb )\;$ respectively as
($\;\partial_\mu\equiv \frac{\partial\;\;}{\partial x^\mu}$)
\bea
\hJ_{\alpha\beta}&=&x_\alpha \;\partial_\beta - x_\beta\; \partial_\alpha = \hX_{\alpha\beta}\cr
\hP_\alpha &=& -\partial_\alpha = \frac{1}{\ro}\;( {\hat p}_\alpha + \hX_{\alpha +})\cr
\hD &=& x^\beta\;\partial_\beta + x^+\;\partial_+ = {\hat p}_+\cr
\hK_\alpha &=& \left( -(\vx^2 + (x^+)^2 )\;\delta_\alpha^\beta +
2\; x_\alpha\; x^\beta\right) \; \partial_\beta + 2\; x_\alpha\;
x^+\;\partial_+ = \ro\;( {\hat p}_\alpha - \hX_{\alpha+})\label{infgencon}
\eea
These operators obeys the $SO(\dm + 1,\dM)$ algebra in the form
\bea
[J_{\alpha\beta}, J_{\gamma\delta}] &=& \eta_{\beta\gamma}\; J_{\alpha\delta} +
\eta_{\alpha\delta}\; J_{\beta\gamma} -(\alpha\leftrightarrow\beta)\cr
[\hP_\gamma , \hJ_{\alpha\beta}] &=& \eta_{\alpha\gamma}\;\hP_\beta -
                                    \eta_{\beta\gamma}\;\hP_\alpha\cr
[\hK_\gamma , \hJ_{\alpha\beta}] &=& \eta_{\alpha\gamma}\;\hK_\beta -
                                    \eta_{\beta\gamma}\;\hK_\alpha\cr
[\hP_\alpha ,\hD ] &=& \hP_\alpha \cr
[\hK_\alpha ,\hD ] &=& -\hK_\alpha \cr
[\hK_\alpha , \hP_\beta] &=& 2\; (\eta_{\alpha\beta}\; \hD +\hJ_{\alpha\beta})\cr
0 &=& [\hP_\alpha ,\hP_\beta ] = [\hK_\alpha , \hK_\beta] = [\hD , \hJ_{\alpha\beta}]
\eea
which is the standard form of the  conformal algebra.
In conclusion, the standard  $SO(\dm +1,\dM)$ linear realization leads to the usual
representation of their infinitesimal generators \eqn{infgen} in the space of functions
on $\min$; the non-linear realization instead yields the representation \eqn{infgenads}
of them in the space of functions on $\ads$, which in turns take the form \eqn{infgencon}
in Poincar\`e coordinates where the conformal algebra point of view is clear, in particular
at the boundary $x^+ = 0$.

\section{Coset representation and geodesics}
\cleqn

$\ads$ spaces can be defined as  pseudo-Riemannian homogeneous symmetric spaces, i.e.
a coset space $G/H$; more precisely: $\; \ads \sim SO(\dm +1 ,\dM)/SO(\dm,\dM)\;$.
This can be seen directly from the coset decomposition introduced in
\eqn{dec1}; let us rewrite the coset element in the form
\bea
K(\vS) &=& \exp \matriz{0}{\vu^t \eta}{\vu}{0} =
\matriz{C}{\vS^t \; \eta}{\vS}{1 + \frac{\vS \; \vS^t\;\eta}{1+C} }\cr
\vS &=& \frac{\sinh \sqrt{\vu^2}}{\sqrt{\vu^2}}\; \vu\cr
-1 &=& - C^2  + \vS^2\label{adscos}
\eea
From here and \eqn{cons} we can identify our defining coordinates as
\bea
w^- &\equiv& \ro\; C\cr
w^\mu &\equiv& \ro\; S^\mu \label{coorads}
\eea
Furthermore it is possible to show \cite{gil} that the metric induced on it
by the invariant metric on $SO(\dm +1,\dM)$ is given by \eqn{gdc}.

There exists an elegant method in this context to get the geodesics wrt this metric \cite{gil}.
It consists simply in lifting up to the coset element the geodesics through the null element
in the Lie algebra wrt the Cartan-Killing metric on it.
In this way we obtain a geodesic through the identity in the coset space that we identify
with the point $(w^{-} = \ro , \vec w =\vec 0)$ in $\ads$.
\footnote{
More exactly, this is so if $\dm=0$ according to the choice of connected component
discussed in Section 2.1.
If $\dm >0$ we have two sets of geodesics  if they are not time-like, one covering the region
$w^- >0$ corresponding to take $w^- = +{\ro}$ as the origin, and one covering the region
$w^- < 0$ if starting through $w^- = -\ro$.
If they  are time-like instead, there is just one family that rolls up the time direction,
see below.
}
In order to get a geodesic through an arbitrary point we have to act by left action with an
element of the coset.
Let us show how all this works here.

First of all let us consider geodesics  through the origin of $\ads$.
They are given by \eqn{adscos} with $\vu$ replaced by
\be
\vec{\bar u}(\tau) = \vu\; \tau\label{geoalg}
\ee
where the ``velocity" $\vu$ parametrizes the family.
This is so because \eqn{geoalg} are the geodesics through the null element in the Lie
algebra.
Therefore according to \eqn{adscos}, \eqn{coorads} the geodesics in $\ads$ through the origin
will be given in $(w^\mu)$-coordinates by
\bea
{\bar w}^\mu (\tau) = \ro\; \left\{ \begin{array}{l}
  \sinh \omega\tau \; {\check u}^\mu \;\; ,\; {\check u}^2 = +1 \;\;,\; space-like\\
\;\;\sin \omega\tau \; {\check u}^\mu \;\; ,\; {\check u}^2 = -1 \;\;,\; time-like\\
\;\tau \;u^\mu\;\;\;\;\;\;\;\;\;\; ,  \;\vu^2 = 0 \;\;\;\;\;,\; null\end{array}\right.
\eea
where $\; \omega = \sqrt{|\vu^2|}\; , {\check u}=\vu /\omega$.
For sake of interpretation is useful to have them in global coordinates $(r_L ,\vnM ,\vnm )$.
If $\vu = (\vu_- , \vu_+)$ and we reintroduce the vector
\be
\vwM \equiv \ro\; \sinh r_L \; \vnM \;\; \in \Re^\dM
\ee
parametrizing the space-like sections of $\ads$, then

\bigskip
\noindent\underline{Space-like geodesics:}
with $\; {\check u}_+{}^2 = 1 + {\check u}_-{}^2 \ge 1$ we have
\bea
\vec{\bar w}_+ (\tau) &=& \ro\; {\check u}_+ \; \sinh \omega\tau\cr
\check{\bar n}_- (\tau) &=& (1 + {\check u}_+{}^2 \sinh^2\omega\tau )^{-\frac{1}{2}}\;
\left( \begin{array}{l} \cosh\omega\tau\\ \sinh\omega\tau\;{\check u}_- \end{array}
\right)\label{geosp}
\eea

\noindent\underline{Time-like geodesics:}
with $\; {\check u}_-{}^2 = 1 + {\check u}_+{}^2 \ge 1$ we have
\bea
\vec{\bar w}_+ (\tau) &=& \ro\; {\check u}_+ \; \sin \omega\tau\cr
\check{\bar n}_- (\tau) &=& (1 + {\check u}_+{}^2 \sin^2\omega\tau )^{-\frac{1}{2}}\;
\left( \begin{array}{l} \cos\omega\tau\\ \sin\omega\tau\;{\check u}_- \end{array} \right)
\label{geoti}
\eea
\noindent\underline{Null geodesics:}
with $\; {\vu}_+{}^2 = {\vu}_-{}^2$ we have
\bea
\vec{\bar w}_+ (\tau) &=& \ro\; {\vu}_+ \;\tau\cr
\check{\bar n}_- (\tau) &=& (1 + {\vu}_+{}^2 \tau^2 )^{-\frac{1}{2}}\;
\left( \begin{array}{c} 1\\ \tau\;{\vu}_- \end{array} \right)\label{geonu}
\eea
From these explicit expressions it is easy for example to compute for a non-null
geodesic the distance between the origin and a point at $\tau$; it is simply:
$\; s(\tau) = \omega\;\ro\;\tau$.
Null geodesics reach the time-like boundary asymptotically in a conserved spatial
direction $\vu_+/\sqrt{\vu_+{}^2}$.
Instead time-like geodesics are bounded and periodic with period $T = \frac{2\pi}{\omega}$;
they never reach the boundary.
\footnote{
For $\dm =1 $, $\;{\check n}_+ = \left(\begin{array}{c} \cos t\\
\sin t\end{array}\right)\;$ with $\; t\sim t+ 2\pi$ and then they wrap the time-like $S^1$
and re-converge to the starting point.
If, as usual made not to have closed time-like loops, $S^1$ is replaced by $\Re$, the
geodesics converge to an image point and diverge from it to re-converge to another one and so on.
}

We complete the analysis given the expression of the geodesics through an arbitrary
point.
From the stated at the beginning of this section  we need to get the map
\be
\vS \longrightarrow {}^{\vS_0}\vS \label{map}
\ee
defined by the left action of an element in the coset in the form
\be
K(\vS_0 )\; K(\vS ) \equiv K( {}^{\vS_0}\vS )\; H(1,Q) \;\;\; , \; Q\in SO(\dm,\dM)
\ee
Then having obtained the map ${}^{\vS_0}\vS$ a geodesic through $\vS_0$ is just
\be
\vec{\bar S} (\tau; \vS_0, \vu ) = {}^{\vS_0}\vec{\bar S}(\tau)
\ee
where $\vec{\bar S}(\tau)$ is a geodesic passing through the origin.
After a little bit of algebra we get the map in the simple form
\be
\left( \begin{array}{l} {\bar C}(\tau; \vS_0 ,\vu) \\ \vec{\bar S}(\tau; \vS_0 ,\vu)\end{array}
\right) = K(\vS_0)\;\left( \begin{array}{l}{\bar C}(\tau)\\ \vec{\bar S}(\tau)\end{array}\right)
\ee
From here, the identifications \eqn{coorads} and the results \eqn{geosp}, \eqn{geoti},
\eqn{geonu}  we have the explicit whole set of geodesics in $\ads$ spaces.

\section{Conclusions}
\cleqn

First of all it is worth to say that we have concentrated along the paper on $\ads$ spaces,
but it should be clear that $\ds$ ones are straigthforwardly studied in our metric
signature convention  starting from \eqn{cons} with a $``+"$ sign on its r.h.s. and working
on $\;{\cal M}_{\dm,\dM+1}\;$.
In particular the coset representation for them is given by $SO(\dm,\dM +1)/SO(\dm,\dM)$.

We obtained explicit global non linear realizations of the isometry groups, equations
\eqn{nltgc}, \eqn{nltpc}, by using various coset decompositions where the parameter space of
the transformations is manifest at difference what happens by considering infinitesimal
transformations.
From these realizations issues like the conformal compactification of the boundary as well as
the relation between standard generators of orthogonal, AdS and conformal algebras are
evident.

Finally it is none to say that geodesics in spaces of physical interest are well-known,
\footnote{
See for example reference \cite{he}, chapter 5.
}
in particular $AdS_{1,3}$.
We believed useful and instructive however to extend these results to arbitrary $\ads$ spaces
giving the explicit expressions obtained through the coset space method without any reference
to the metric.

We think our results can be of usefulness among other things in the non trivial problem of
fixing boundary conditions in the search of solutions of gravity-matter classical field theories
in the presence of a non zero cosmological constant \cite{ls}.

\section*{Acknowledments} We thank to Fidel Schaposnik  for discussions and references.

\end{document}